\documentclass[twocolumn]{aastex62}

\graphicspath{{./}{figures/}}

\received{January 1, 2018}
\revised{January 7, 2018}
\accepted{\today}
\submitjournal{ApJ}
\shorttitle{Sulfur Abundance in Slow Solar Winds}
\shortauthors{Kuroda \& Laming}

\begin{document}

\title{Magnetic Field Geometry and Composition Variation in Slow Solar Winds: The Case of Sulfur}

\correspondingauthor{Natsuha Kuroda}
\email{natsuha.kuroda.ctr.ja@nrl.navy.mil, nkuroda@ucar.edu}

\author[0000-0001-9260-8555]{Natsuha Kuroda}
\affil{University Corporation for Atmospheric Research \\
PO Box 3000 \\
Boulder, CO 80307-3000, USA}
\affil{Space Science Division, Code 7684, Naval Research Laboratory, Washington DC 20375, USA}

\author[0000-0002-3362-7040]{J. Martin Laming}
\affil{Space Science Division, Code 7684, Naval Research Laboratory, Washington DC 20375, USA}

\begin{abstract}

We present an examination of the First Ionization Potential (FIP)
fractionation scenario invoking the ponderomotive force in the chromosphere, and its
implications for the source(s) of slow speed solar winds by
using observations from  \textit{The Advanced Composition Explorer}
(\textit{ACE}). Following a recent conjecture that
the abundance enhancements of intermediate FIP elements, S, P, and C, in
slow solar winds can be explained by the release of plasma fractionated on open fields, though from
regions of stronger magnetic field than usually associated with fast solar
wind source regions, we identify a period in 2008 containing four solar
rotation cycles that show repeated pattern of sulfur abundance enhancement
corresponding to a decrease in solar wind speed. We identify the source
regions of these slow winds in global magnetic field models and find that
they lie at the boundaries between a coronal hole and its adjacent active
region, with origins in both closed and open initial field configurations. Based on magnetic field extrapolations, we model the fractionation and
compare our results with element abundances measured by \textit{ACE} to estimate the
solar wind contributions from open and closed field, and to highlight potentially useful directions for further work.

\end{abstract}

\keywords{solar wind -- Sun: abundances -- Sun: chromosphere -- turbulence -- waves}

\section{Introduction} \label{sec:intro}

The First Ionization Potential (FIP) fractionation effect, the phenomenon
where abundances of low FIP elements (FIP $<$ 10 eV) are observed to be
higher in corona than in the photosphere compared to high FIP elements (FIP $>$
10 eV), has been studied for over five decades
\citep{2004ApJ...614.1063L,1985ApJS...57..151M,1985ApJS...57..173M,1963ApJ...137..945P,2012ApJ...755...33S}.
It is known that the fractionation pattern in the solar wind can be generally
categorized by solar wind type, and therefore understanding the fractionation
mechanism can help us study the properties of the source regions of various
solar winds. The fast winds ($\gtrsim$ 600 km/s) are known to show a low
degree of FIP fractionation \citep{2007A&ARv..14....1B,1998ApJ...505..999F}
and are associated with open field lines emanating from coronal holes. On the
other hand, slow winds ($\lesssim$ 600 km/s) from quiet sun and active
regions show a high degree of FIP fractionation
\citep{1998ApJ...505..999F,2012ApJ...755...33S}. Spectroscopic observations
of open and closed field show similar variations in FIP fractionation,
supporting the origin of fast winds in coronal holes, and suggesting that slow
winds are associated with closed loop structures on the solar surface which are
subsequently opened up by reconnection with neighboring open field lines, a
process known as interchange reconnection, though slow winds coming directly from open field is not ruled out.

While there are several theoretical models that attempt to explain the FIP
effect
\citep{1994AdSpR..14..139A,1998SoPh..182..293A,1999ApJ...521..859S,1989A&A...225..222V},
\citet{2004ApJ...614.1063L} first introduced a model invoking the
ponderomotive force, which arises from the interaction between chromospheric
plasma and Alfv\'{e}n waves propagating through or reflecting from the
chromosphere. This model can reproduce the observed fractionation pattern
from both open and closed field configurations \citep{2015LRSP...12....2L},
and also provides insight into the Inverse FIP effect
\citep{2019ApJ...875...35B}. Recently, more subtle variations in FIP
fractionation have been noted. \citet{2018SoPh..293...47R} pointed out
that intermediate FIP elements such as S, P, and C show a higher degree of
fractionation in energetic particles originating from co-rotating interaction
regions (CIRs) than they do in particles collected during gradual solar
energetic particle events (SEPs), and that similar abundance
pattern variations are seen in the presumed source regions for these
energetic particle samples; slow speed solar winds for the particles from
CIRs \citep[see e.g.][]{2007AJ....134.2451G,2013ApJ...776...92K} and the
closed loop solar corona for SEPs \citep[e.g.][]{1995ApJ...443..416L}. \citet[][see Fig. 2 therein]{2020SSRv..216...20R} uses these and other abundance anomalies to develop models of the provenance of various SEP populations.

\citet{2019arXiv190509319L} investigate this further, and find for a closed
loop supporting resonant Alfv\'en waves (where the wave travel time from one
loop footpoint to the other is an integral number of half wave periods)
insignificant fractionation of S, P and C, while for open field (where no
Alfv\'en wave resonance exists) S, P and C can become fractionated if the
magnetic field is sufficiently strong, much stronger than that usually
associated with fast solar wind origins. Therefore these abundances might offer
some discrimination between slow solar wind originating in open or closed field configurations. 

In this study, we aim to test this hypothesis by examining the possible
correspondence between the enhanced fractionation of intermediate FIP
elements in slow solar winds and the magnetic field geometry of their source
regions which may evolve through interchange reconnection. We do so by surveying
the extensive record of solar wind speeds and composition from \textit{ACE}
mission and investigating magnetic features on the Sun associated with the
abundance anomaly we identify in slow winds. We also obtain profiles of the
magnetic field strength at these source regions, estimate the fractionation
values of various elements using the model outlined in
\citet{2019arXiv190509319L} with the field profiles as inputs, and compare
the results with the observed fractionation values. Section
\ref{sec:observation} introduces the finding from \textit{ACE} data and the
identified magnetic features on the Sun, Section \ref{sec:modeling} discusses
the modeling results, Section \ref{sec:discussion} gives some discussion and
Section \ref{sec:conclusions} concludes.

\section{Reduction of \textit{ACE} data and the identification of slow solar wind source location}
\label{sec:observation}

In this study, we use the \textit{Solar Wind Ion Composition Spectrometer}
\citep[\textit{SWICS};][]{1998SSRv...86..497G} 1.1 data and \textit{The Solar
Wind Electron, Proton, and Alpha Monitor} (\textit{SWEPAM}) data that are
publicly available from \textit{ACE} level 2 database
\footnote{\url{http://www.srl.caltech.edu/ACE/ASC/level2/index.html}}.
The \textit{SWICS} data consists of elemental abundances, charge state compositions,
and kinetic properties of all major solar-wind ions from H through Fe from
launch to August 23, 2011, at time resolutions of 1 hour, 2 hour, and 1 day,
depending on the observable. We took the solar wind speed from the 2-hour data
of He$^{2+}$ bulk velocity, the abundances of He, C, Ne, Mg, Si, Fe with respect
to O from the 2-hour data, and the abundances of N and S with respect to O from
the 1-day data. We averaged all 2-hour data over a day centered at the recorded
time of 1-day data, while excluding the data points that do not have the quality
flag of 0 (``good quality data'', see release note
\footnote{\url{http://www.srl.caltech.edu/ACE/ASC/DATA/level2/ssv4/swics_lv2_v4_release_notes.txt}})
or are associated with the solar wind type of coronal mass ejection (as compared
with streamer winds or coronal hole winds, a rough classification based on functions
of O$^{7+}$/O$^{6+}$ charge state ratio versus proton speed, according to the data description
\footnote{\url{http://www.srl.caltech.edu/ACE/ASC/level2/ssv4_l2desc.html}}).
We then calculated the fractionations and their error values for each element
by dividing the measured abundances and their error values by their respective
photospheric abundances with respect to oxygen; the photospheric abundances
for O, S, C, N, Fe were taken from \citet{2011SoPh..268..255C} and those for Ne,
He, Mg, Si were taken from \citet{2009ARA&A..47..481A}.

Following our aim described in the introduction, we plot the time profiles of
these daily solar wind speeds and elemental fractionation values and visually
survey the record for the expected fractionation pattern in slow solar winds.
We control the search with three qualitative requirements. First, we restrict
our search to the time frame near solar minimum so that the following
identification of the source solar feature of slow winds is less complicated.
Second, we focus on the fractionation enhancement of sulfur, since there are
no phosphorus data and the variation in carbon fractionation values, as we
find, is relatively small and therefore its comparison with wind speeds is
more uncertain. Lastly, we only accept the sulfur enhancements in slow winds
that appear within repeated wind speed patterns over a few months, so that
the identified correspondence is more likely associated
with the same solar features reappearing on the Sun.

\begin{figure*}[ht!]
\plotone{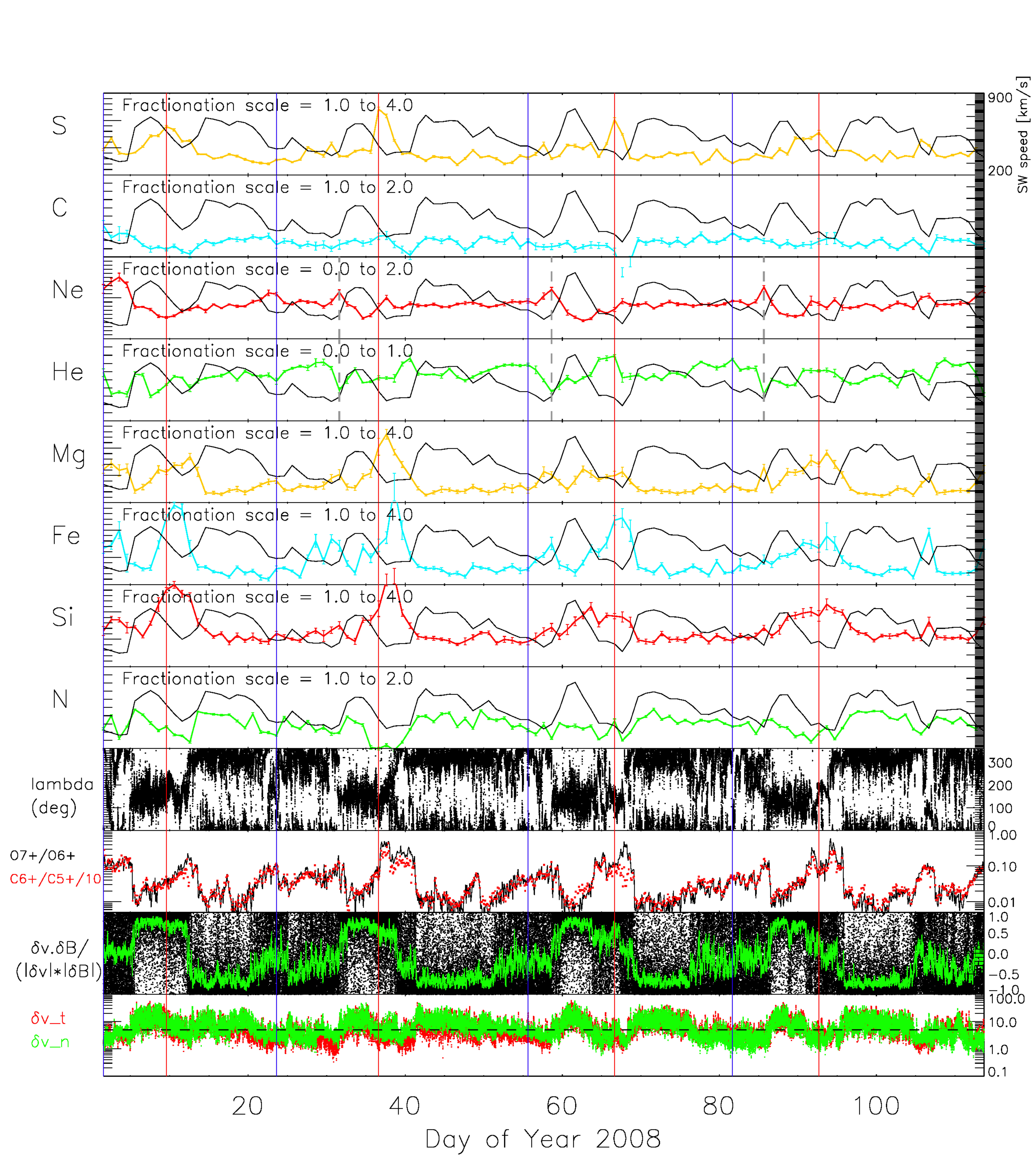}
\caption{Top 8 panels: The daily solar wind speed (black) vs. fractionation values of various elements (colors) from 2008 January 1 to 2008 April 22. Repeated sulfur enhancement within a certain wind speed pattern, as well as interesting correspondences among various elements (see Section \ref{sec:observation} for detail) can be seen. Bottom 4 panels: the orientation of IMF fields ($\sim135$ degrees and $\sim315$ degrees correspond to the inward and outward direction with respect to the nominal Parker spiral, respectively, 64s average), O$^{+7}$/O$^{+6}$ (black) and C$^{+6}$/C$^{+5}/10$ (red) charge state ratios (2-hour average), the quantity $(\delta v \cdot \delta B)/|\delta v||\delta B|$, which is a proxy of Alfv\'{e}nicity (black for 64s average and green for their 1-hour average), and the velocity fluctuations in normal (green) and tangential (red) direction (64s average). The red and blue vertical lines mark the two kinds of slow wind that repeatedly appear during this period, one with sulfur enhancement and the other without sulfur enhancement, respectively.}
\label{fig:tprof_zoom}
\end{figure*}

Figure \ref{fig:tprof_zoom} shows our data from 2008 January 1 to 2008 April 22, around
the minimum of Solar Cycle 23. Each elemental fractionation value is plotted
on an equal scale of max/min = 4 (colors) on top of the solar wind speed (200
km/s to 900 km/s, black). It shows
a repeated sulfur fractionation enhancement corresponding to a drop in the
solar wind speed. The appearance seems to be a part of a repeated solar wind
speed pattern that shows two distinct fast streams, one short-lived followed
by another one that goes over more gradual decrease, separated by the drop.
Although the sulfur enhancement corresponding to the solar wind speed drop is
present before and after this period too, the level of increase with respect
to the low-level fractionation during surrounding times is quite distinct
during this period. There are several other interesting features to note
during this period. Low-FIP elements, Mg, Fe, and Si, all show enhanced
fractionation with the sulfur enhancements.  Based on observed abundance differences between coronal holes and closed magnetic loops, it had been long suspected (possibly naively) that
the low-FIP elements fractionate more in winds originating in closed field
structures, so their enhancements would suggest the repeated appearance of
such structures on the solar surface. However S does not fractionate in this
way in closed loops \citep{1995ApJ...443..416L,2019arXiv190509319L}, and a high S/O
abundance ratio might point to a solar wind origin in open field, a point we explore in more detail below. 

Neon during this period shows a very distinct fractionation variation; the lowest
fractionation value corresponding with the first fast wind, relatively sharp
increase corresponding to the following slow wind, then steady values over
mild decrease in the solar wind speed, followed by an uptick to the highest
value at the second minimum of the wind speed cycle.
\citet{2014ApJ...789...60S} studied the solar wind neon abundance and its
dependence on wind speed and evolution with the solar cycle using the same
data as this study but in a statistical manner over the entire mission
period. They find that the neon abundance values can be categorized by three
solar wind types: fast wind from coronal holes which has the lowest neon/oxygen
abundance ratio of Ne/O$\sim0.097\pm0.014$, slow wind from active regions which
prevail during solar maximum with slightly higher abundance ratio of
Ne/O$\sim0.116\pm0.017$, and slow wind from quiet sun (helmet streamers)
which prevail during solar minimum with the highest abundance ratio of Ne/O
$\sim0.170\pm0.025$ \citep{2014ApJ...789...60S}. 

In this work, the abundance ratio values for
each phase we identify in this period corresponds with the results of \citet{2014ApJ...789...60S}:
$\sim0.074$ for the first fast wind, $\sim0.126$ for the first slow wind,
then $\sim0.163$ for the second slow wind at the end of cycle. This Ne/O abundance
ratio enhancement in the second slow wind coincides with a depletion in the He/O abundance ratio  at DoY 32, 58 and 85 (marked with grey vertical dashed lines in Figure \ref{fig:tprof_zoom}),
a circumstance identified with gravitational settling in a coronal loop prior to
release into the solar wind \citep{2019arXiv190509319L}. Otherwise, the helium variation
shows noticeable enhancements corresponding to sulfur enhancements during
this time period (for the first, the third, and the last red vertical lines in Figure \ref{fig:tprof_zoom}, most pronounced at the third one on DOY 66), which we return to in Section 4.

\begin{figure*}[ht!]
\plotone{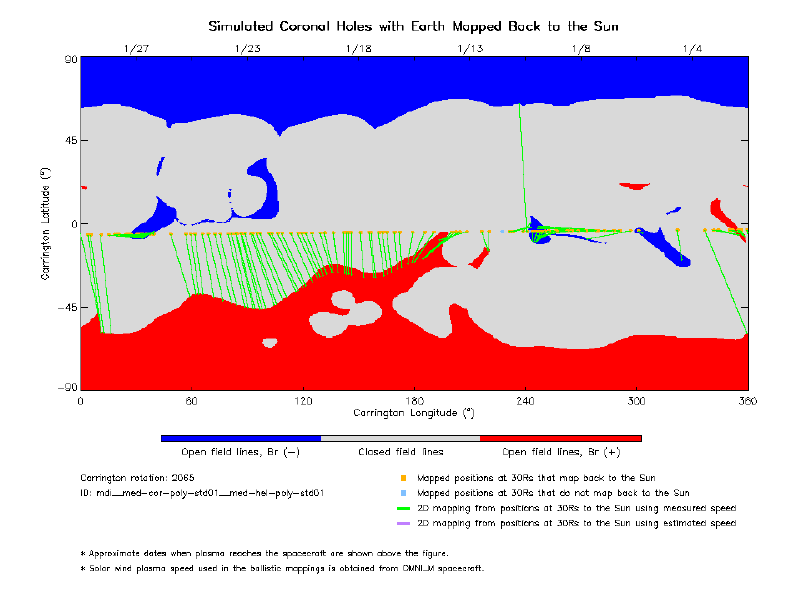}
\plotone{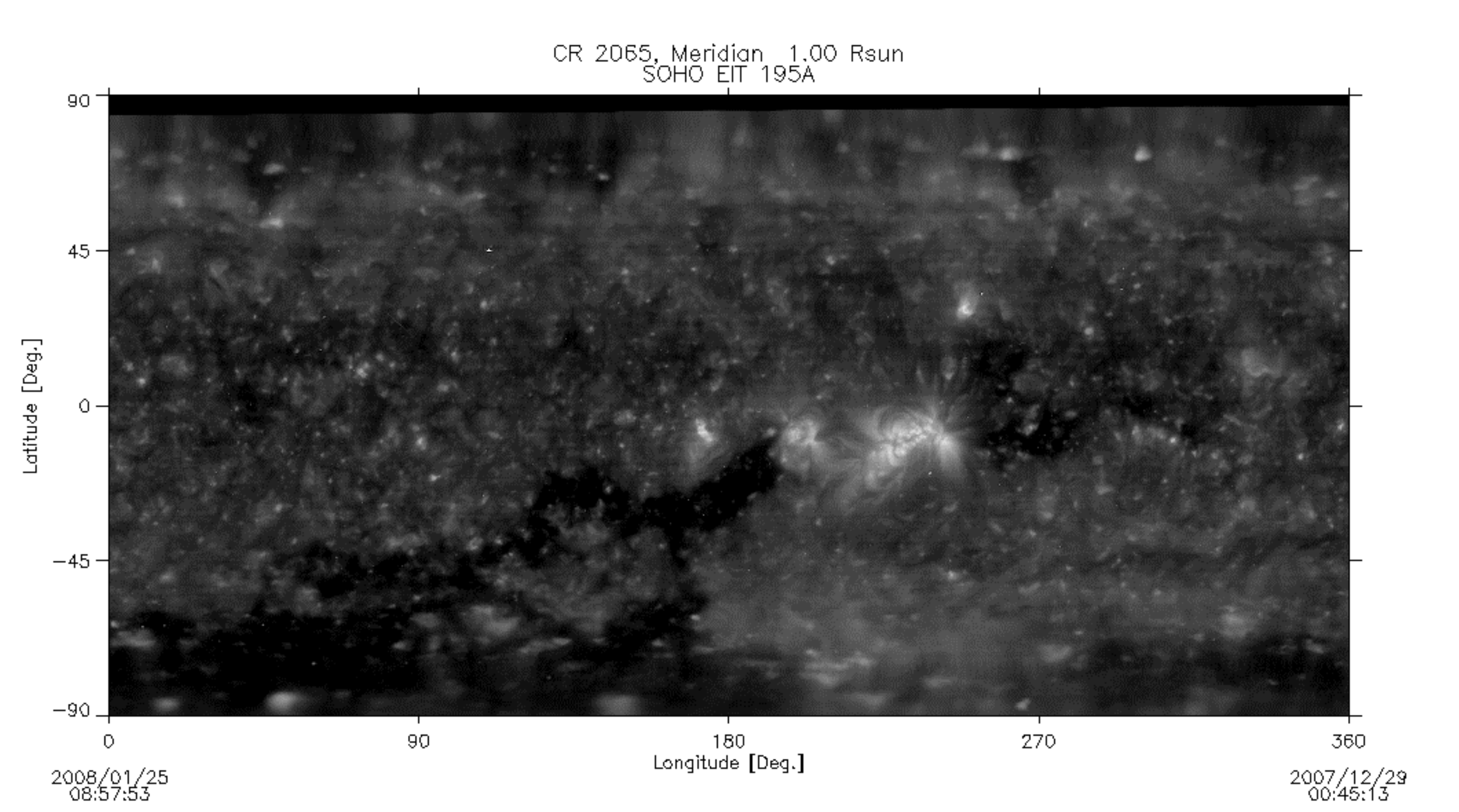}
\caption{(a) Spacecraft mapping of the solar wind plasma at 1 AU, available from
\url{http://www.predsci.com/\%20mhdweb/spacecraft_mapping.php} for the first
Carrington rotation cycle (2065). The users are to locate the yellow circle corresponding to the target date and follow the green line emanating from it to identify the source location of the observed solar wind back on the Sun. The first sulfur-enhanced slow winds were observed over January 4 to January 11, which suggests their source locations to be the negative coronal hole located at $\sim 250$ degrees, followed by the adjacent AR 10980. (b) The corresponding synoptic map from 195 \AA{} channel of the \textit{EIT} on board \textit{SOHO}.}
\label{fig:synoptic}
\end{figure*}

\begin{figure*}[ht!]
\epsscale{1.2}
\plotone{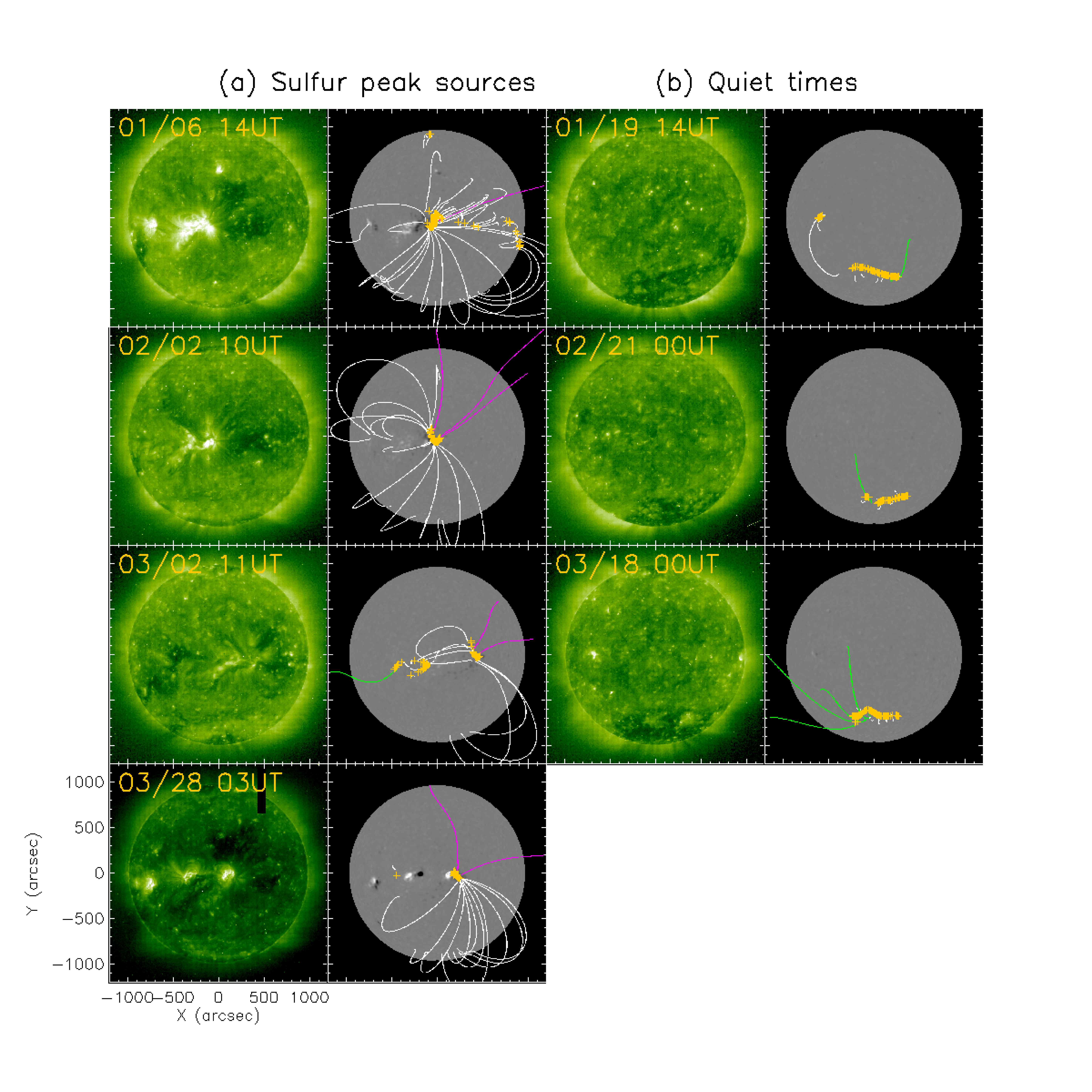}
\caption{Field lines identified at the footpoint locations (orange crosses on the magnetograms) of the sulfur-enhanced slow wind ((a)) and of the regular slow wind ((b)), marked by the red and blue vertical lines in Figure \ref{fig:tprof_zoom}, respectively, with corresponding  \textit{EIT} 195\AA{} images. The white, magenta, and green field lines indicate the closed, negative open, and positive open field lines, respectively. The average magnetic field profiles of open field lines (only negative/magenta for (a)) were used for the modeling in Section \ref{sec:modeling}.}
\label{fig:sun_images}
\end{figure*}

Figure \ref{fig:synoptic} shows the spacecraft mapping data of the solar
wind plasma at 1 AU ((a), available from
\url{http://www.predsci.com/\%20mhdweb/spacecraft_mapping.php}) and the
corresponding synoptic map from 195 \AA {} channel of the \textit{Extreme
ultraviolet Imaging Telescope} (\textit{EIT}) \citep{1995SoPh..162..291D} on board \textit{The Solar and
Heliospheric Observatory} (\textit{SOHO}) \citep{1995SoPh..162....1D} for the
first Carrington rotation cycle (CR2065) during this time period. For (a),
the users are to locate the yellow circle corresponding to the target date
and follow the green line emanating from it to identify the source location
of the observed solar wind back on the Sun. If one identifies the sources of
the winds detected on the dates over which the contained sulfur level rises
and decays in CR2065 (from January 4 to January 11), it is apparent that the
majority points to the equatorial coronal hole at the Carrington
longitude$\sim$250 degrees with negative polarization, which is adjacent to
an active region (AR10980). We confirmed that this pair of the negative
equatorial coronal hole and the adjacent active region consistently appears
at every sulfur peaks during this period (the AR 10980 decays over three
rotations and the new AR 10987 appears at the same location in the fourth
rotation, see Figure \ref{fig:sun_images} (a)). Together with the low-FIP
(Mg, Fe, Si) observation, it seems that the repeated sulfur enhancement in
the first slow wind is possibly related to this coronal hole-active region
pair structure.


We additionally show in Figure \ref{fig:tprof_zoom} some of the data
from \textit{SWEPAM} during the same period in the bottom four panels. The first, lambda, is the
64s-average interplanetary (IP) magnetic field longitude in RTN coordinates,
which indicates the magnetic field direction with respect to the nominal
Parker spiral (inward and outward for $\sim135$ degrees and $\sim315$
degrees, respectively). The second is the hourly-averaged
O$^{+7}/$O$^{+6}$ (black) and C$^{+6}/$C$^{+5}/10$ (red) charge state ratios.
The third is the quantity $(\delta v \cdot \delta B)/|\delta v||\delta
B|$, which is the normalized quantity that measures the degree of correlation
between the IP velocity and magnetic fluctuation. This can be used as a proxy
of Alfv\'{e}nicity \citep{1987JGR....9211021R,2018ApJ...864..139K}, which
measures how much of the fluctuations in the solar wind stream consist of
pure Alfv\'{e}n waves. Values of +1 or -1 mean pure Alfv\'{e}n waves traveling  in the opposite
or the same direction as the mean field direction, respectively. The last
is the velocity fluctuation in radial (r) and tangential (t) dimension.
\citet{2018ApJ...864..139K} found that many dynamic parameters including the
ones plotted in Figure \ref{fig:tprof_zoom} show distinctive changes in
tandem during the interval defined by the low level of velocity fluctuations
($\delta v < 5$ km/s, which is marked by the black dashed horizontal line in
the bottom panel), which roughly coincides with interval between two fast
wind streams. They found that, during this interval, Alfv\'{e}nicity becomes
closer to 0, the charge state ratio increases, and the IMF lambda
reverses/does not change for the streams associated with a heliospheric
current sheet (HCS)/pseudostreamer (PS) crossing where the two adjacent wind
streams are from magnetic sectors of opposite/same polarity. They argued that
this is due to the nature of the slow wind being more like the ``boundary
layer'' from which it is generated as the certain spatial structure between
two streams passes across the solar surface. We see in Figure
\ref{fig:tprof_zoom} that there are two kinds of slow wind in every cycle in
terms of the above-mentioned characteristics found by
\citet{2018ApJ...864..139K}, and one of them is associated with sulfur
enhancement (marked by red vertical lines) and the other one is not (marked
representatively by blue vertical lines based on the Alfv\'{e}nicity value).
The former is clearly associated with IMF lambda reversal from inward to
outward in two streams, which corresponds with two coronal holes of opposite
sign located over Carrington longitude of $\sim$200--250 degrees in Figure
\ref{fig:synoptic} (a), adjacent to the active region. On the other hand, the
latter is associated with no sign change of IMF direction, and its outward
direction corresponds to the large positive coronal hole shown in Carrington
longitude of $\sim$0--180 degrees in Figure \ref{fig:synoptic} (a), with no
active region nearby. We therefore believe that our repeated sulfur
enhancement is most likely related to the spatial structure
of coronal hole-active region pair on solar surface.



\begin{figure*}[ht!]
\plotone{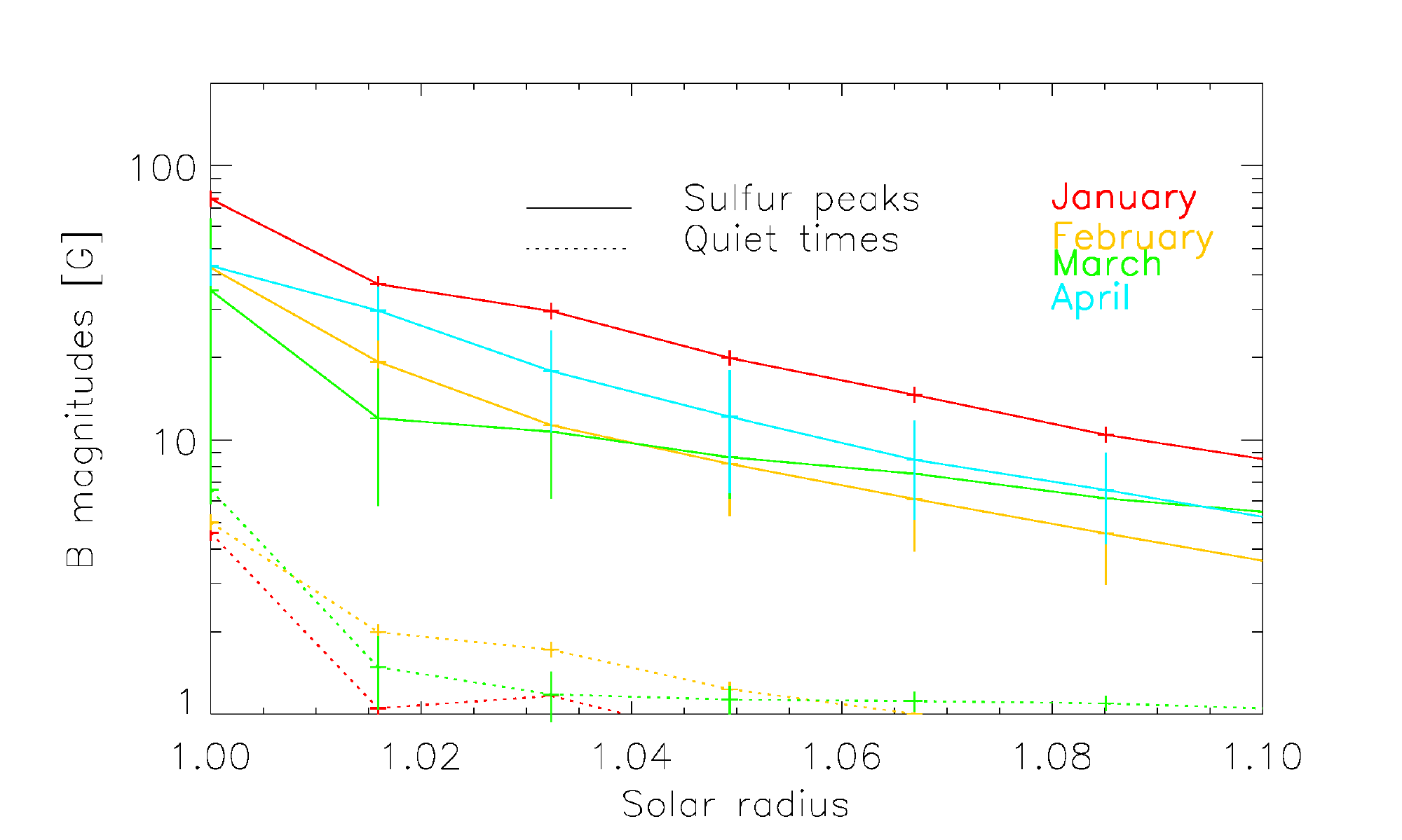}
\caption{The average magnetic field profiles for the open field lines identified
at the source regions of sulfur-enhanced solar winds and regular solar winds.
The fields are stronger for the former because they all emanate from the region near active regions.}
\label{fig:field_profile}
\end{figure*}

Next, we investigated the magnetic field geometry of the source region of the sulfur-enhanced slow winds. To do so, we
used the 1-hour cadence spacecraft mapping data of the solar wind plasma at 1 AU shown in
Figure \ref{fig:synoptic} (a) and the Potential Field Source Surface
\citep[pfss, 6-hour cadence;][]{2003SoPh..212..165S} package available in SolarSoft \citep{1998SoPh..182..497F}.
First, we obtain from the spacecraft mapping data all possible footpoint locations of the solar wind plasma
detected at 1 AU over several days (4--7 days) covering the sulfur peak day (marked with red vertical lines in
Figure \ref{fig:tprof_zoom}). The beginning and the end date of this date range is defined by the first local
minima of sulfur fractionation values counting backward and forward from the peak day. Next, we estimate the
time that plasma detected on the sulfur peak day left the sun by dividing the Sun-spacecraft distance on the
sulfur peak day by the solar wind speed measured on the sulfur peak day
(both available from the spacecraft mapping data). Then, we obtain the pfss model closest to this time, and extract all field
lines originating closest to the identified possible footpoint locations.
Figure \ref{fig:sun_images} (a) shows the resulting field lines next to the corresponding \textit{EIT} 195\AA{} images.
It is apparent that sulfur-enhanced slow winds all come from the structure containing open fields from coronal hole and
closed fields that belong to the adjacent active region. We also show the results for the ``regular''
slow winds marked by the blue vertical lines in Figure \ref{fig:tprof_zoom}.
The date range over which the solar wind footpoint locations were extracted was 4 days for all three dates.
Compared to the sulfur peak dates, the quiet time solar wind traces back to the open fields with no significant magnetic feature nearby. Figure \ref{fig:field_profile} shows the average magnetic field strength profile of all identified open field lines shown in Figure \ref{fig:sun_images} for each day. Note that, for March sulfur peak, the positive open field line seen in Figure \ref{fig:sun_images} was not included in the averaging process because the suggested IMF direction in Figure \ref{fig:tprof_zoom} for this date was inward. It is apparent that the field strengths are higher for the source regions of sulfur-enhanced slow winds since these field lines all originate near active regions. In the next section, we will use this magnetic field profile as an input to the ponderomotive force model of the FIP fractionation and compare the resultant fractionation of various elements with the observation.

\section{FIP modeling}
\label{sec:modeling} In this section we attempt a quantitative interpretation
of the abundance variations based on the model of fractionation by the
ponderomotive force, guided by the magnetic field reconstructions. Alfv\'en
waves causing the FIP fractionation can have either a coronal origin,
presumably excited by nanoflares in closed loops \citep{2016ApJ...831..160D}, or a
photospheric origin deriving ultimately from fluctuations excited by
convective motions. As in previous work \citep[e.g.][]{2015LRSP...12....2L} we construct
model magnetic fields for open and closed field cases, and using the empirical  
chromospheric model of \citet{avrett08} integrate the
Alfv\'en wave transport equations for a full non-Wentzel-Kramers-Brillouin
(non-WKB) treatment of the wave propagation. For the time being we ignore
terms giving wave damping or growth, and the wave origin only comes about as
a matter of interpretation.

In closed loops, we take waves from nanoflares to be excited at the loop
resonance or its harmonics. This is in general a higher frequency than the
waves associated with convective motions originating lower down in the
atmosphere, and the solution gives largest wave amplitude in the corona and
ponderomotive force concentrated at the top of the chromosphere. In open
loops, we take waves of period five minutes deriving from convection. Once
the non-WKB wave solution is found by integrating the Alfv\'en wave transport
equations, the instantaneous ponderomotive acceleration, $a$,
acting on an ion is evaluated from the general form \citep[see e.g. the appendix
of][]{2017ApJ...844..153L}
\begin{equation}
a={c^2\over 2}{\partial\over\partial z}\left(\delta E^2\over B^2\right)
\end{equation}
where $\delta E$ is the wave (transverse) electric field, $B$ the ambient (longitudinal) magnetic field,
$c$ the speed of light, and $z$ is a coordinate along the magnetic field.

Given the ponderomotive acceleration, element fractionation is calculated
using input from the chromospheric model and the equation
\citep{2017ApJ...844..153L}
\begin{eqnarray}
\nonumber f_k&=&{\rho _k\left(z_u\right)\over\rho _k\left(z_l\right)}\\ &=&\exp\left\{
\int _{z_l}^{z_u}{2\xi _ka\nu _{kn}/\left[\xi _k\nu
_{kn} +\left(1-\xi _k\right)\nu _{ki}\right]\over 2k_{\rm B}T/m_k+v_{||,osc}^2+2u_k^2}dz\right\}.
\end{eqnarray}
This equation is derived from the momentum equations for ions and neutrals in
a background of protons and neutral hydrogen. Here $\xi _k$ is the element
ionization fraction, $\nu _{ki}$ and $\nu _{kn}$ are collision frequencies of
ions and neutrals with the background gas \citep[mainly hydrogen and protons,
given by formulae in][]{2004ApJ...614.1063L}, $k_{\rm B}T/m_k \left(
=v_z^2\right)$ represents the square of the element thermal velocity along
the $z$-direction, $u_k$ is the upward flow speed and $v_{||,osc}$ a
longitudinal oscillatory speed, corresponding to upward and downward
propagating sound waves. At the top of the
chromosphere where background H is becoming ionized $\nu _{ki}>>\nu _{kn}$,
and small departures of $\xi _k$ from unity can result in significant
decreases in the fractionation. This feature is important in inhibiting the
abundance enhancements of S, P, and C at the top of the chromosphere. Lower down where
the H is neutral this inequality does not hold, and these elements can
become fractionated.

The longitudinal oscillatory speed $v_{||,osc}$ is composed of sound waves
deriving from convection \citep[see][for the most recent
implementation]{2019arXiv190509319L} and sound waves excited by the Alfv\'en
wave driver added in quadrature \citep[see][]{2017ApJ...844..153L}. When this
quadrature sum exceeds the local Alfv\'en speed, we assume that the resulting
shock will produce sufficient turbulence and mixing to completely restrict
further fractionation. For a parallel magneto-hydrodynamic shock, the first
critical Alfv\'en Mach number is unity, and the shock must generate
turbulence. Some observational evidence is given by
\citet{2008ApJ...683L.207R}.

In all prior work \citep[e.g.]{2015LRSP...12....2L, 2017ApJ...844..153L,
2019arXiv190509319L} we have restricted the region of fractionation to be
above the chromospheric equipartition layer where sound speed and Alfv\'en
speed are equal. All fractionation is assumed to occur in low plasma-$\beta$
gas, where the plasma-$\beta = 8\pi nk_{\rm B}T/B^2$ is the ratio of gas pressure to magnetic pressure.
In the closed field geometry supporting resonant waves, this assumption
makes no difference, because the ponderomotive acceleration is restricted to
the top of the chromosphere by the non-WKB solution in any case. Here we give
some further justification for this assumption in the open field situation.

\begin{table*}
\begin{center}
\caption{Model Fractionations for Slow Solar Wind Epochs 0, 2, 4 and 6}
\begin{tabular}{l|cc|cc|cc|cc}
\hline
ratio & \multicolumn{2}{c}{0} & \multicolumn{2}{c}{2}& \multicolumn{2}{c}{4}& \multicolumn{2}{c}{6} \\
& open& closed&  open& closed&  open& closed&  open& closed\\
\hline
Mg/O& 2.92& 2.62& 3.36& 2.87& 2.84& 2.66& 2.89& 2.51\\
Fe/O& 2.47& 2.53&  2.87& 2.83& 2.76& 2.75& 2.46& 2.45\\
Si/O&  2.72&  1.98& 3.01& 2.14& 2.42& 2.04& 2.63& 1.94\\
S/O&  2.26& 1.28&  2.29& 1.34& 1.72& 1.33& 2.06& 1.29\\
C/O&  2.37&  1.10& 2.33& 1.12& 1.50& 1.09& 2.14& 1.11\\
N/O&  1.01& 0.78&  0.99& 0.78& 0.87& 0.78& 1.00& 0.80\\
Ne/O& 0.94& 0.67& 0.93& 0.67& 0.80& 0.68& 0.94& 0.70\\
He/O& 1.16& 0.55& 1.10& 0.54& 0.73& 0.54& 1.13& 0.57\\

\hline
\end{tabular}
\end{center}
\end{table*}

The dominant wave mode in the $\beta > 1$ part of the atmosphere is acoustic,
which can propagate at all angles to the magnetic field, and can effectively
cascade to microscopic scales to cause mixing of the plasma, effectively
quenching any fractionation. Higher up, where $\beta < 1$, the magnetic field
structures the plasma. Magnetosonic waves, which can propagate across the
magnetic field, can escape laterally, leaving the Alfv\'en and acoustic waves, both of which
are constrained to travel close to the magnetic field direction, in the FIP fractionation
region. This constraint inhibits
their cascade and the consequent plasma mixing, allowing the ponderomotive
force to fractionate the plasma.  The waves driven
in the $\beta > 1$ region are taken to be kink waves in magnetic flux
concentrations \citep[e.g.][]{2005ApJS..156..265C, 2013A&A...554A.115S,
2015A&A...577A..17S}. \citet{2005ApJS..156..265C} describe how kink modes in
the structured atmosphere at $\beta > 1$ evolve to become transverse Alfv\'en
waves as the magnetic field expands to fill up the volume when $\beta < 1$.
The mechanism of fractionation depends on the
interaction of waves and ions (but not neutrals) through the refractive index
of the plasma, and the effects this has on the refraction and reflection of
waves. Waves are reflected (in the case of Alfv\'en waves) or refracted (for
fast modes), and the resulting change in momentum of the wave is balanced by
an impulse on the ions \citep[see e.g.][for an optical analog]{ashkin70}.

In addition
to the fractionation coming from the ponderomotive force, a mass dependent fractionation
comes from conservation of the first adiabatic invariant. This is evaluated from the magnetic
field line expansion between close to the solar surface (see below) and 1.5 solar radii heliocentric distance. In
\citet{2019arXiv190509319L} we argued that at this point, the solar wind becomes collisionless,
in the sense that the solar wind speed divided by the heliocentric radius becomes larger than
the ion-proton collision frequency, $\nu _{ip}$, the reasoning being that at this point diffusion can no longer 
supply particles from the solar disk to counteract the abundance deficit caused by the extra
acceleration, $a$. In Appendix A we give a slightly more rigorous treatment. The abundance modifications ``freeze-in''
when the ion thermal velocity $v_t < a/\nu _{ip}$, which leads to the same numerical conclusion.

\begin{table}
\begin{center}
\caption{Model Fractionations for Slow Solar Wind Epochs 1, 3 and 5}
\begin{tabular}{l|cc|cc|cc}
\hline
ratio & \multicolumn{2}{c}{1} & \multicolumn{2}{c}{3}& \multicolumn{2}{c}{5} \\
& open& closed&  open& closed&  open& closed\\
\hline
Mg/O& 1.73& 1.61& 2.14& 1.93& 1.70& 1.55\\
Fe/O& 1.68& 1.69& 1.97&  1.97& 1.61& 1.56\\
Si/O& 1.72&  1.42& 2.12&  1.61& 1.68& 1.37\\
S/O&  1.59&  1.14& 1.90&  1.19& 1.56& 1.10\\
C/O&  1.40& 0.98&  1.62&  0.98& 1.40& 0.98\\
N/O&  1.06& 0.82&  1.08&  0.78& 1.06& 0.83\\
Ne/O& 0.92& 0.76& 0.90&  0.69& 0.92& 0.75\\
He/O& 0.81& 0.61& 0.76&  0.51& 0.82& 0.65\\
\hline
\end{tabular}
\end{center}
\end{table}

\begin{table}
\begin{center}
\caption{Magnetic Field Parameters for Slow Solar Wind Epochs}
\begin{tabular}{l|ccccccc}
\hline
epoch & 0&1&2&3&4&5&6 \\
\hline
$B\left(1.015 R_{\sun}\right)$ (G)& 37&1.05&19.3&2.0&12.0&1.5&29.7\\
$B\left(1.5 R_{\sun}\right) (G)$&0.47&0.32&0.26&0.12&0.59&0.25&0.57\\
$\ln\left(B_{1.015}/B_{1.5}\right)$ &4.4&1.2&4.3&2.8&3.0&1.8&3.9\\
\hline
\end{tabular}
\end{center}
\end{table}

\begin{figure*}[ht!]
\epsscale{1.2}
\plotone{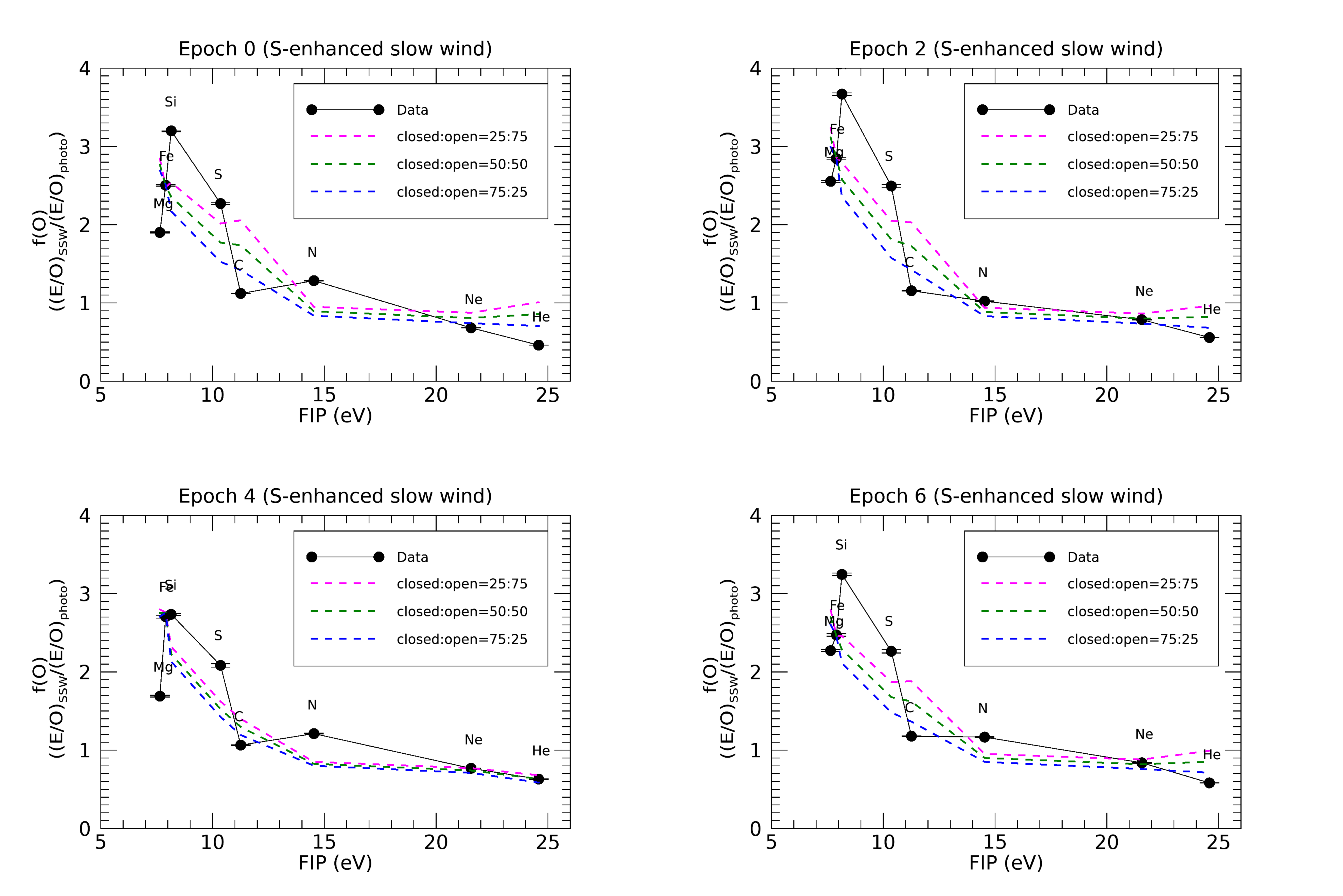}
\caption{Fractionations with respect to O for Mg, Fe, Si, S, N, Ne, and He, plotted for
epochs 0, 2, 4, 6 (black data points). Model fractionations give different proportions of
plasma fractionated in open and closed field. Open field favors high S/O and high He/O.}
\label{fig:epoch0}
\end{figure*}

The PFSS extrapolations use photospheric magnetograms as the lower boundary condition for the calculation of the coronal magnetic field.
No account is taken here of magnetic field line expansion through the chromosphere as the plasma transitions from
being gas pressure dominated ($\beta > 1$) to being magnetic pressure dominated ($\beta < 1$).
We estimate the magnetic field in the ($\beta < 1$) region of the chromosphere for the open field models from the values given by the
extrapolations at 1.015 R$_{\sun}$ (see Figure \ref{fig:field_profile}). Model fractionations for open and closed field cases are given Tables 1 
and 2, for the
``sulfur enhanced'' slow wind (epochs 0, 2, 4, 6) and for epochs 1, 3, and 5 without strong sulfur fractionation, respectively, based on the magnetic field parameters given in Table 3.
For the closed field calculations we assume 30 G and 2 G in Tables 1 and 2 respectively, though the fractionations are not sensitive to these values.
Waves on closed loops are assumed to be shear Alfv\'en waves with amplitude in the corona around 30 - 100 km s$^{-1}$, depending
on the coronal density assumed, while those on open fields are assumed to be torsional. The amplitude of the torsional wave varies depending
on the magnetic field, with lower wave amplitudes ($\sim 25$ km s$^{-1}$ at 1.7 R$_{\sun}$ where the integrations back to the Sun are 
started) required on the higher magnetic field epochs in Table 1
to provide the fractionation, rising to $\sim 200$ km s$^{-1}$ for epoch 4. This is because with higher magnetic field, fractionation
occurs across a wider range of altitudes in the chromosphere, and corresponding smaller ponderomotive force is required.
In Table 2, where the open magnetic field is much smaller, wave amplitudes at 1.7 R$_{\sun}$ of 400 - 500 km s$^{-1}$ are required.
For reference, the polar coronal hole model of \citet{2005ApJS..156..265C} gives (presumably r.m.s., derived from line broadening observations) 
wave amplitudes in this region of 100 - 200 km s$^{-1}$, and the fast solar wind models in \citet{2019arXiv190509319L} require 300 km s$^{-1}$. The open field regions considered here have significantly lower magnetic fields
than a typical polar coronal hole, and so might reasonably be expected to have higher wave amplitudes due to refraction effects.

For these open low magnetic field models, we have made an extra change to the chromospheric model. 
The hydrogen ionization balance given by \citet{avrett08}, upon which 
our models are based, is elevated over that that would result from thermal equilibrium. This is presumably attributed to the effect of
chromospheric shock waves, as in e.g. \citet{carlsson02}. A better match of theory to observations is found for the low open magnetic field regions by enforcing
thermal equilibrium in the chromosphere, and using this assumption to calculate the explicit hydrogen ionization balance. The increased neutral fraction of hydrogen increases the
degree of fractionation of other elements for a given ponderomotive force, and increases the relative fractionation of S/O and C/O. We argue that such ``quiescent''
chromosphere, with low magnetic fields and relatively faint in emission lines would not be picked up by the empirical procedures of
\citet{avrett08}, but nevertheless should be expected at the footpoints of open low magnetic field regions, where heating is taken to be proportional
to magnetic field strength \citep[e.g.][]{oran17}, and insignificant heat is conducted back downwards into the chromosphere from higher altitudes.


\section{Discussion}
\label{sec:discussion}

\begin{figure*}[ht!]
\epsscale{1.2}
\plotone{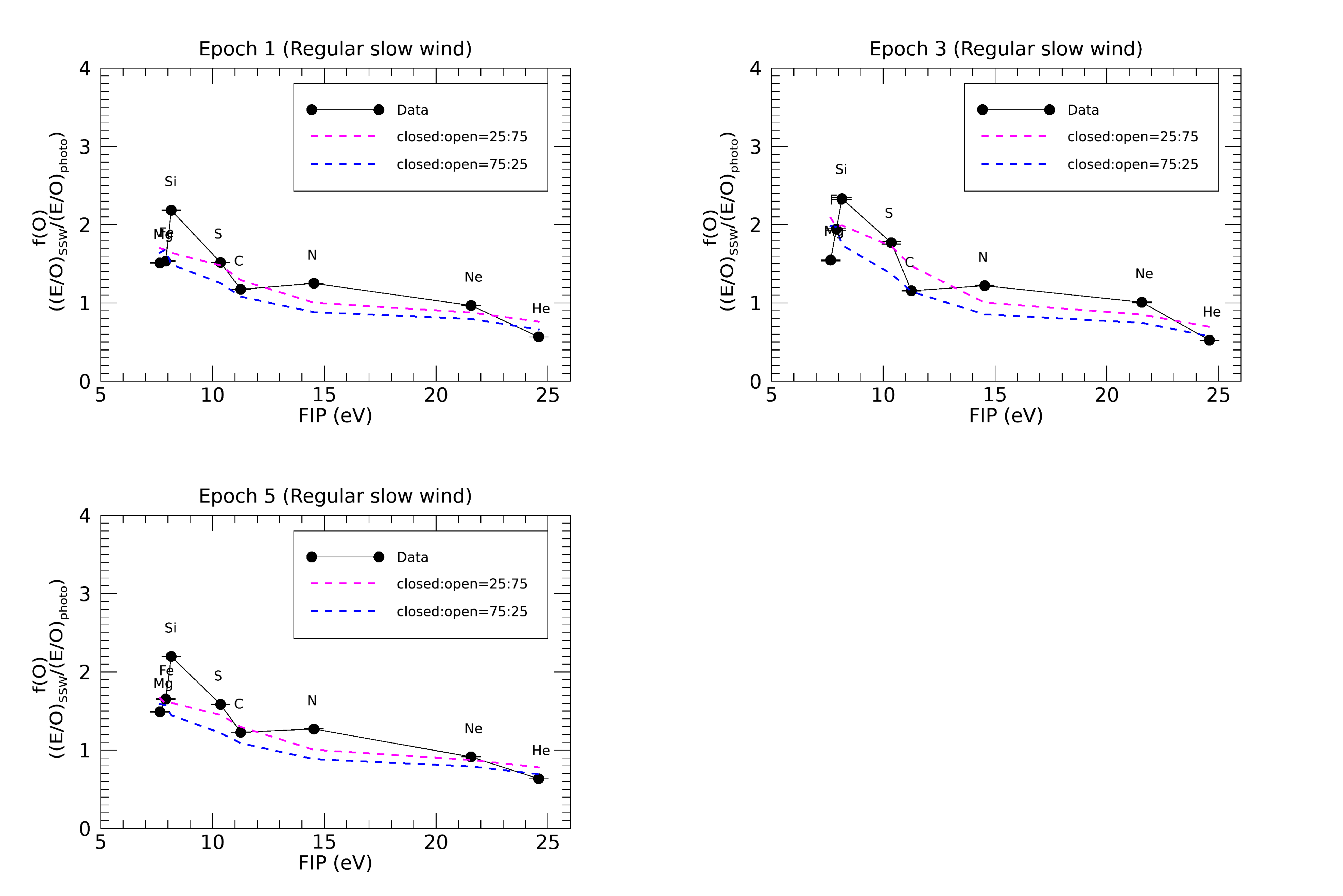}
\caption{Fractionations with respect to O for Mg, Fe, Si, S, N, Ne, and He, plotted for
epochs 1, 3, 5 (black data points). Model fractionations give different proportions of
plasma fractionated in open and closed field.}
\label{fig:epoch1}
\end{figure*}

The observed fractionations of various elements for each day are taken by
averaging the values over the date ranges used for the solar wind footpoint
identification in Section \ref{sec:observation}. 
In Figure \ref{fig:epoch0} we compare data from the slow wind corresponding to the enhanced S/O abundance
ratio with models, highlighted by the red lines in Figure \ref{fig:tprof_zoom}, plotted as black symbols and compared to model fractionations calculated for various mixtures of plasma from closed and open magnetic field regions. The data are taken from
approximately six day intervals, encompassing the peaks in the S/O, Mg/O, Fe/O and Si/O abundance ratios. The
models for closed and open fields are designed to match the observed Fe/O ratio, with open field models reproducing 
the observed enhancement of S/O. Epochs 0, 2, 4 and 6 show qualitatively similar behavior to each other,
with the S abundance relative to Fe being highest for the highest magnetic fields, (0, 6, 2, and 4 in descending order), both in data
and in the open field models. The models also show a greater difference between open and closed field geometries for higher $B$. This is
to be expected, since as explained above, the closed field only fractionates at the top of the chromosphere, while fractionation
can occur from the $\beta = 1$ layer upwards in open field, and the $\beta =1$ layer is pushed to lower altitudes when $B$
is higher. In general the open field models enhance S/O and C/O, while the closed field does not. The closed field depletes He/O 
while the open field does not, though Ne/O is well reproduced by both models. Roughly equal amounts of plasma fractionated
in closed and open field appear to be present in this sample of slow speed solar wind.

The fractionation of Mg and Si with respect to Fe is not predicted by the models, and has been seen previously \citep{pilleri15}.
Curiously, \citet{heidrich18} re-analyze the same data using a different numerical procedure and find a different fractionation. For
the range of O$^{7+}$/O$^{6+}$ charge state ratios (0.04 - 0.1) determined from Figure \ref{fig:tprof_zoom} 
(around the red vertical lines), \citet{heidrich18} find Mg fractionated slightly more than Fe, and both are fractionated 
more than Si. This is the reverse of the trend found here and by \citet{pilleri15}, and is in better agreement with our models.
While the fractionation of Fe relative to Mg or Si is affected by the mass dependency introduced by the conservation of the
first adiabatic invariant, the relative fractionations of Mg and Si depend much more on the ponderomotive force, with Mg
being about 10\% greater than Si in open field, and about 30\% greater in closed loops.

Neither \citet{pilleri15} nor \citet{heidrich18} considered the fractionation of S. Our models at least qualitatively capture this effect,
with S considerably more fractionated in the open field than in the closed loop. Further support for this idea comes from a detailed
inspection of Figure \ref{fig:tprof_zoom}. The strong peaks in 
S/O appear {\em before} the corresponding peaks in Mg, Fe, and Si relative to O, by 0.5 -1 days, and both of these sets of features
arrive at \textit{ACE} about 0.5 days ahead of the low Alfv\'enicity turbulence. Bearing in mind that the minor ions generally flow faster
than the bulk solar wind by a significant fraction of the Alfv\'en speed, $ \alpha v_A$, and that the low Alfv\'enicity turbulence will be
advected with the bulk solar wind, we interpret the strong peaks in Mg/O,  Fe/O and Si/O as coming from a closed field region
associated with the low-Alfv\'enicity turbulence, but arriving in advance at the spacecraft because of their faster speeds. In Appendix
B we estimate this time difference to be $\sim \alpha v_A/v_s$ times the bulk solar wind travel time where $v_s$ is the bulk
solar wind speed. This evaluates to a time of order 0.5 days, as observed. We further interpret the S/O peak to be more associated
with the open field region and high Alfv\'enicity period just before the closed loop plasma, although at this point there will be considerable
overlap between plasma originally in closed and originally in open fields.

C/O is also not observed to be enhanced along with S/O, as predicted by our models, and also based on expectations
from prior work on SEPs \citep{2018SoPh..293...47R}, and solar wind measured by \textit{Ulysses} \citep{vonsteiger00} 
and \textit{ACE} \citep{reisenfeld07}. These works typically give the C/O abundance ratio around 0.7, representing a fractionation of about 1.3 based on photospheric abundance of \citet{2011SoPh..268..255C},
though recent measurements of samples returned by the Genesis mission \citep{heber13,laming+17} give a ratio closer to 1, or a fractionation
of $\sim 1.8$.
For reasons that remain obscure, the C/O ratio measured herein on both open and closed fields stays close to that expected from closed field fractionation, with values 1 - 1.2, only
marginally consistent with results quoted above.
We have tried estimating the effect of molecular CO on the C and O fractionation. Taking partition functions from \citet{rossi85}
and assuming local thermodynamic equilibrium we calculate the fraction of C and O atoms bound in molecules, and modify the fractionation
calculations accordingly. In practice, outside of sunspots, this is a negligible effect. Similarly, the formation of H$_2$ to increase the neutral 
fraction of the background gas is not a significant factor. 

It is worth noting the possible difference caused by different versions of \textit{ACE} level-2 data; \textit{SWICS} 1.1 data underwent a major new release around March 2015 using completely redesigned analysis method based on \citet{2014ApJ...789...60S}. We evaluated the possible effect that this change had in our study by recreating Figure 3 of \citet{reisenfeld07}, which used the older version of \textit{ACE}/\textit{SWICS} data. We find that C/O fractionation is lower in the new data because O abundance is higher in the new data. We also find that Mg/Fe is less than 1 in the new data, which is in line with the discrepancy we mentioned earlier (Si/Fe is also less than 1 but higher than Mg/Fe). Despite these discrepancies, we find S/Fe to be consistent in both data sets: $\sim$0.78 in the new data and $\sim$0.74 in the old data. Therefore, the sulfur trend we find in this study seems to be unaffected by this data version change.

In Figure \ref{fig:epoch1} we show similar plots for the epochs of non-S enhanced slow wind, which from Figure \ref{fig:field_profile}
is associated with much lower magnetic field strengths. The difference in fractionation between open and closed field is now smaller, as
the $\beta = 1$ layer moves upwards in the chromosphere, closer for open field to the region where fractionation is restricted to by the closed
field. There is now better agreement in general between models and fractionations, (though fractionations are lower). The C/S ratio
prefers the thermal equilibrium chromosphere discussed above, otherwise C is strongly overpredicted. This is the case for epochs 0, 2, 
4, and 6,
where the approximation of thermal equilibrium is less plausible here than for epochs 1-3-5, due to the higher magnetic field and
increased heating \citep[e.g.][]{oran17}. Among the low FIP ions,  Si/O is consistently underpredicted, while the high FIP elements N/O, Ne/O and
He/O are reproduced well. There are smaller differences here between open and closed field models due to the small magnetic field.

Several previous authors have studied high-Alfv\'enicity slow speed solar
winds, either from the point of view of magnetic fields and
waves/turbulence \citep[e.g.][]{bale19, wang19} together with imaging \citep{rouillard20}, or also including
considerations of the wind composition \citep[e.g.][]{damicis19, owens20,
stansby20}. These last two make comparisons with the He/H abundance ratio,
with \citet{stansby20} finding He/H observed by {\it Helios} \citep{porsche77} in Alfv\'enic slow speed solar winds
comparable to that in fast winds, and significantly higher than that found
in non Alfv\'enic slow winds. This is in qualitative agreement with our results in Tables 1 and 2, though the precise quantitative agreement is less clear because {\it ACE/SWICS} and our models specify He/O. One feature we address that these other authors do not is the
dependence of the He fractionation on the strength of the chromospheric magnetic field, which also seems to be borne out by our dataset.

Finally, for completeness, we mention solar wind periods shortly after epochs 3 and 5 where the Ne/O abundance ratio is enhanced,
while the He/O abundance ratio is depleted. This is argued in \citet{2019arXiv190509319L} to be a signature of pre-release gravitational 
settling, as found in \textit{ACE}/\textit{SWICS} observations by \citet{weberg12,weberg15}. Here, heavy elements are seen to be depleted with respect to H.
The Ne/O and He/O behavior arises because He is most affected by gravitational settling, followed by O, and then Ne. Figure 6 
in \citet{weberg12} indicates that they do indeed detect this region as a heavy ion dropout.

\section{Conclusions}
\label{sec:conclusions}
We have investigated a period of solar wind showing various repeatable element abundance modifications with a view to understanding
the extent to which these can be explained by the ponderomotive force model of the ion-neutral separation (the
FIP Effect), and testing some of the assumptions embedded in the model. Of the high FIP elements, S, P, and C have the lowest FIPs,
and can be significantly ionized in the chromosphere. When the background gas is also significantly ionized, these elements do not fractionate 
due to back diffusion of the neutral fraction; only true low FIP elements fractionate. But when the background gas is neutral, minor elements 
ions and neutrals move through the background gas with equal ease, and S, P, and C can become fractionated. In our models, this can occur 
in open field regions where the magnetic field is sufficiently strong, due to the assumption that FIP fractionation only occurs above the plasma 
$\beta =1$ layer (discussed in Section 3). This arises because the $\beta =1$ layer is pushed to lower and lower altitudes by the increasing 
 magnetic field, allowing more of the fractionation to occur in chromosphere plasma where H is neutral, consequently increasing S, P, and C. 
This effect is restricted to open fields, on the assumption that waves on closed loops are dominated by waves that are resonant with the loop, 
in the sense that the wave travel time from one footpoint to the other is an integral number of wave half-periods. In this case, the wave 
solution restricts fractionation to the upper chromosphere, no matter what the magnetic field strength is, and S, P, and C remain relatively 
unfractionated.

The thrust of this paper has been to study to what extent these ideas can be validated by direct solar wind observations. We have found 
solar wind intervals with large S/O abundance ratios that do indeed appear to be coming from open field regions with high magnetic field, 
and other solar wind periods associated with low magnetic field open region without significant S/O fractionation. While these conclusions largely
depend on PFSS magnetic field extrapolations to identify solar wind source regions, they are also supported by observations of the
Alfv\'enicity of the solar wind, and especially in the case of the large S/O solar wind, the time delay between the S/O peak and those for other low FIP elements, Fe/O, Mg/O, Si/O, etc, which probably come from closed field associated with low Alfv\'enicity solar wind. We suspect that the accuracy of our abundance modeling is mainly limited by the chromospheric model. \citet{avrett08} provide a static 1D ``average'' empirical chromosphere, which is certainly adequate for establishing the validity of the ponderomotive force as the agent behind the fractionation, and for understanding long term averages of solar wind abundances. More accurate fractionations will probably require incorporation of more details of chromospheric dynamics \citep[e.g.][]{carlsson02,carlsson19}.

\acknowledgements This research is supported by the NASA Living With a Star Jack Eddy Postdoctoral Fellowship Program, administered by UCAR's Cooperative Programs for the Advancement of Earth System Science (CPAESS) under award NNX16AK22G, by grants from the NASA
Heliophysics Supporting Research (NNH16AC39I), Heliophysics Grand Challenges (NNH17AE96I) and the
Laboratory Analysis of Returned Samples Programs (NNH17AE60I), and by basic
research funds of the Chief of Naval Research. JML also acknowledges the
hospitality of the International Space Science Institute in Bern where some
of this work was started.

\appendix
\section{Abundance Modification by Adiabatic Invariant Conservation}
The transport equation for the solar wind distribution function $f$ is written
\begin{equation}
{\partial f\over\partial t} +{\bf v}\cdot\nabla f +{{\bf F}\over m}\cdot\nabla _{\bf v} f
= {\rm scattering~terms} + {\rm particle~source~terms}
\end{equation}
with $f\propto \exp -\left({\bf v}-{\bf U}\right)^2/2v_t^2$
where ${\bf U}$ is the solar wind bulk velocity, ${\bf v}$ are the velocities of solar wind particles, and $v_t$ is the thermal speed.
The force ${\bf F}$ includes all forces. In steady-state conditions, with ${\bf U} = U{\bf \hat{r}}$,
${\bf F} = F{\bf \hat{r}}$
and ${\bf B} = B{\bf \hat{r}}$ where ${\bf \hat{r}}$ is a unit vector in the radial direction, and separating
out the action of the first adiabatic invariant from ${\bf F}$ we get,
\begin{equation}
v_{\|}{\partial f\over\partial r}+{v_{\|}v_{\perp}\over 2}{\nabla _{\|}B\over B}
{\partial f\over\partial v_{\perp}} + \left[{F\over m}-{v_{\perp}^2\over 2}
{\nabla _{\|}B\over B}\right]{\partial f\over\partial v_{\|}} = \nu _{ip}\left(v_t\over v\right)^3{\partial\over\partial\mu }
\left[\left(1-\mu ^2\right){\partial f\over\partial\mu}\right],
\end{equation}
where a term for minor ion collisions with protons has been included on the right hand side.
With $\partial f/\partial v_{\|}=-\left(v_{\|}-U\right)/v_t^2$ and $\partial f/\partial v_{\perp}=-v_{\perp}/v_t^2$,
and averaging over $v_{\|}$ and $v_{\perp}$,
\begin{equation}
U{\partial \left<f\right>\over\partial r}= {U\over 2}{\nabla _{\|}B\over B}\left<f\right>
= -{a\over v_t^2}\left<f\right>.
\end{equation}
where $a=v_t^2\nabla _{\|}B/2B$ is the acceleration resulting from the first adiabatic invariant conservation. In conditions where $\nabla _{\|}B/B < 0$, the particle velocity increases and if
nothing else happens, the abundance decreases to maintain constant particle flux. However diffusion induced by the negative
concentration gradient will increase particle fluxes and abundances in the solar wind. The net flux will then be
\begin{equation}
J = -D{\partial\left<f\right>\over\partial r} = {\rm min}\left(v_t, -{v_t^2\over\nu _{ip}}{\nabla _{\|}B\over B}\right)\left<f\right>,
\end{equation}
where the diffusion coefficient $D = v_t^2/\nu _{ii}$, and the diffusive flux is limited to flow at less than the particle thermal speed. This
implies a ``freeze-in'' of abundances when $\nu _{ip} < -v_t\nabla _{\|}B/B$. Compared with equation 1 in \citet{2019arXiv190509319L},
the term on the right hand side has changed from $v_{sw}/r$, where $v_{sw}$ is the solar wind velocity and $r$ the heliocentric radius, to $v_t/\left|l_B\right|$, where $l_B$ is the scale length of ${\bf B}$. Numerically, these amount to the same conclusion; abundances freeze-in at
a density of $10^5 - 10^6$ cm$^{-3}$ at a heliocentric radius of $\sim 1.5 R_{\sun}$.

\section{Parker Spiral Travel Times of Minor Ions and Background Plasma}
Minor ions are known to flow at some fraction of the Alfv\'en speed, $\alpha v_A$, where $\alpha \sim 0.5$, faster than the bulk
plasma, and consequently will arrive at {\it in situ} detectors before the plasma and entrained waves with which they might be associated.
We consider a simple model where the solar wind speed, $v_s$, and the Alfv\'en speed, $v_A$ are taken as constant. The background plasma travel time is then
\begin{equation}
t_{plasma} = \int _{R_{\sun}}^r {dr\over v_s} = {r-R_{\sun}\over v_s}
\end{equation}
while the minor ion travel time is
\begin{equation}
t_{ions} = \int _{R_{\sun}}^r {dr\over v_s+\alpha v_A\cos\left(\phi - \phi _0\right)}\simeq
 \int _{R_{\sun}}^r {dr\over v_s+\alpha v_A\cos\left({\omega\sin\theta\over v_s}\left(r-R_{\sun}\right)\right)}
\end{equation}
where the Parker spiral is given by the approximate equation ${\omega\sin\theta\over v_s}\left(r-R_{\sun}-R_{\sun}
\ln{r\over R_{\sun}}\right)= \phi - \phi _0$ \citep[e.g][]{priest14}, $\omega = 2.8\times 10^{-6}$ rad s$^{-1}$ being the
angular velocity of the solar rotation, $\theta$ the polar angle, with $\theta =\pi/2$ in the ecliptic plane,and  $\phi$ and $\phi _0$
are the azimuthal angle made by the field line to the radial direction, and its initial value at $r=R_{\sun}$. An approximation
to this equation has been substituted into the final form in equation B6. We write
$z= {\omega\sin\theta\over v_s}\left(r-R_{\sun}-R_{\sun}\right)$ and integrate to find
\begin{equation}
t_{ions} = {2v_s\over\omega\sin\theta \sqrt{v_s^2-\alpha^2v_A^2}}\arctan\left\{\left(v_s-\alpha v_A\right)\tan{\omega\sin\theta
\over 2v_s}\left(r-R_{\sun}\right)\over  \sqrt{v_s^2-\alpha^2v_A^2}\right\}
\end{equation}
where $v_s > v_A$. If $v_A\rightarrow 0$, $t_{ions} \rightarrow \left(r-R_{\sun}\right)/v_s$ as expected. For $v_A>0$ but $<< v_s$,
we find an approximate expression
\begin{equation}
t_{ions} \simeq {r-R_{\sun}\over v_s}\left(1 - {\alpha v_A\over v_s} + \cdots\right)  \simeq t_{plasma} -  {r-R_{\sun}\over v_s}
{\alpha v_A\over v_s}.
\end{equation}
Thus minor ions ($\alpha\sim 0.5$) will arrive {\em before} low Alfv\'enicity waves  ($\alpha\sim 0$) which are essentially entrained in the
plasma, but {\em after} waves with high Alfv\'enicity, travelling along the field at the Alfv\'en speed  ($\alpha\sim 1$).


\end{document}